\documentclass[a4paper,11pt]{amsart}
\begin{document}
\hyphenation{gra-vi-ta-tio-nal re-la-ti-vi-ty Gaus-sian
re-fe-ren-ce re-la-ti-ve gra-vi-ta-tion Schwarz-schild
ac-cor-dingly gra-vi-ta-tio-nal-ly re-la-ti-vi-stic pro-du-cing
de-ri-va-ti-ve ge-ne-ral ex-pli-citly des-cri-bed ma-the-ma-ti-cal
de-si-gnan-do-si coe-ren-za pro-blem Lam-pa-riel-lo}
\title[Einstein, Levi-Civita, and Bianchi relations]
{{\bf  Einstein, Levi-Civita, and Bianchi relations}}

\author[Angelo Loinger]{Angelo Loinger}
\address{Dipartimento di Fisica, Universit\`a di Milano, Via
Celoria, 16 - 20133 Milano (Italy)}
\email{angelo.loinger@mi.infn.it}
\thanks{To be published on \emph{Spacetime \& Substance.}}

\begin{abstract}
About an essential suggestion to Einstein by Levi-Civita in 1915.
\emph{Unicuique suum}.
\end{abstract}

\maketitle


\vskip1.20cm \noindent In a recent article by Ebner on the history
of general relativity \cite{1} the author remarks (see p.7 and
p.10 of \cite{1}) that in the famous paper of November 25th, 1915
\cite{2}, in which Einstein wrote for the first time the correct
equations of general relativity

\begin{equation} \label{eq:one}
R_{ab}= -\kappa (T_{ab} - \frac{1}{2}\:g_{ab}\:T) \quad, \quad
(a,b=1,2,3,4) \quad \textrm{--}
\end{equation}

equivalently:

\begin{equation} \label{eq:onebis}
R_{ab} - \frac{1}{2}\:g_{ab}\:R = -\kappa \: T_{ab}  \quad
\textrm{--} \quad , \tag{1bis}
\end{equation}

no explanation is given about how he had found the term $(\kappa
/2)g_{ab}T$ -- equivalently, $-(1/2)g_{ab}R$ of (\ref{eq:onebis})
--, which was missing in the previous (November 4th, 1915
\cite{3}) and incomplete version of the theory, i.e. $R_{ab}= -
\kappa T_{ab}$. Now, the contemporary Italian relativists (in
particular, A. Palatini and R. Serini) were aware that the above
term had been suggested to Einstein by Levi-Civita, who had
pointed out that, by virtue of Bianchi identities, the covariant
divergence of $R_{ab}-(1/2)g_{ab}R$ is equal to zero, thus
assuring the compatibility with the differential conservation
relations (and gravitational equations of motion of matter):
covariant divergence of $T_{ab}$ equal to zero.

\par I wish to emphasize that \emph{there exists a written testimony of this
fact}. Indeed, in 1942 prof. Giovanni Lampariello (1903-1964), a
former pupil and a colleague of Tullio Levi-Civita (1873-1941),
wrote a comment on a Levi-Civita's scientific autobiography of
1938, see \cite{4}. After mentioning the fundamental geometric
results of Beltrami and Ricci-Curbastro, the manuscript of
Lampariello specifies the decisive -- mathematically and
physically -- contributions by Levi-Civita to general relativity.
And \emph{in primis} Lampariello recalls ``$\ldots$ l'aggiunta al
tensore gravitazionale di Einstein $[$ i.e. $R_{ab}]$ del termine
correttivo $[$ i.e. $-(1/2)g_{ab}R]$ che lo riduce a divergenza
identicamente nulla, cos\`i da soddisfare alle leggi della
dinamica dei continui $\ldots$'' -- Translation: ``$\ldots$ the
addition to Einstein's gravitational tensor $[$ i.e. $R_{ab}]$ of
the  corrective term $[$ i.e. $-(1/2)g_{ab}R]$ which reduces it to
an identically zero divergence, thus satisfying the dynamical laws
of continuous media $\ldots$'' --

\par Levi-Civita did not give publicity on scientific journals, or
on newspapers, to his epistolary suggestion to Einstein. A superb
instance of unselfishness.\nolinebreak --

\par I am grateful to my friend Dr. S. Antoci, who sent me a copy
of Ebner's essay \cite{1}.

\end{document}